\newcommand{\be}{\begin{equation}}
\newcommand{\ee}{\end{equation}}
\newcommand{\brr}{\begin{eqnarray}}
\newcommand{\err}{\end{eqnarray}}

\newcommand{\bd}{\begin{displaymath}}
\newcommand{\ed}{\end{displaymath}}

\newcommand{\bfig}{\begin{figure}}
\newcommand{\efig}{\end{figure}}

\def\lb{\label}

\documentclass{elsart}
\usepackage{ifvtex}
\usepackage{ifpdf}
\usepackage[dvips]{graphicx}
\usepackage{amssymb}
\usepackage{upgreek}
\usepackage{yfonts}
\usepackage{dsfont}
\journal{Physica A}
\begin{document}
%%%%%%%%%%%%%%%%%%%%%%%%%%%%%%%%%%%%%%%%%%%%%%%%%%%%%%%%%%%%%%%%%%%%%%%%%%%%%%%%%%%%%%%%%%%%%%%%%%%%%%%%%%%%%%%%%%%%%%%%%
\begin{frontmatter}
\title{Spin tunneling in magnetic molecules: Quantitative estimates for Fe8 clusters}
\author{D. Galetti\corauthref{cor}} and 
\corauth[cor]{Corresponding author.}
\ead{galetti@ift.unesp.br}
\author{Evandro C. Silva}
\address{Instituto de F\'{\i}sica Te\'{o}rica, S\~{a}o Paulo State University, 
         Rua Pamplona 145, 01405-900, S\~{a}o Paulo, SP, Brazil}
%%%%%%%%%%%%%%%%%%%%%%%%%%%%%%%%%%%%%%%%%%%%%%%%%%%%%%%%%%%%%%%%%%%%%%%%%%%%%%%%%%%%%%%%%%%%%%%%%%%%%%%%%%%%%%%%%%%%%%%%%
\begin{abstract}
Spin tunneling in the particular case of the magnetic molecular cluster
octanuclear iron(III), Fe8, is treated by an effective Hamiltonian that
allows for an angle-based description of the process. The presence of an
external magnetic field along the easy axis is also taken into account in
this description. \ Analytic expressions for the energy levels and barriers
are obtained from a harmonic approximation of the potential function which
give results in good agreement with the experimental results. The energy
splittings due to spin tunneling is treated in an adapted WKB approach and
it is shown that the present description can give results to a reliable
degree of accuracy.
\end{abstract}
%%%%%%%%%%%%%%%%%%%%%%%%%%%%%%%%%%%%%%%%%%%%%%%%%%%%%%%%%%%%%%%%%%%%%%%%%%%%%%%%%%%%%%%%%%%%%%%%%%%%%%%%%%%%%%%%%%%%%%%%%
%%%%%%%%%%%%%%%%%%%%%%%%%%%%%%%%%%%%%%%%%%%%%%%%%%%%%%%%%%%%%%%%%%%%%%%%%%%%%%%%%%%%%%%%%%%%%%%%%%%%%%%%%%%%%%%%%%%%%%%%%
\end{frontmatter}
%%%%%%%%%%%%%%%%%%%%%%%%%%%%%%%%%%%%%%%%%%%%%%%%%%%%%%%%%%%%%%%%%%%%%%%%%%%%%%%%%%%%%%%%%%%%%%%%%%%%%%%%%%%%%%%%%%%%%%%%%
\section{Introduction}
%%%%%%%%%%%%%%%%%%%%%%%%%%%%%%%%%%%%%%%%%%%%%%%%%%%%%%%%%%%%%%%%%%%%%%%%%%%%%%%%%%%%%%%%%%%%%%%%%%%%%%%%%%%%%%%%%%%%%%%%%

In the last decades, tunneling in mesoscopic systems has attracted a great
deal of interest \cite{takagi,chudteja,legget}. This physical process
corresponds in a standard quantum description to the tunneling of the
collective degree of freedom associated with the angular momentum direction
through a potential barrier separating two minima of an effective potential
associated with the spatial orientation. Besides the interest in this new
class of processes due to the wide domain of investigations and possible
applications, it also draws attention because it can shed light on several
aspects of our understanding of the transition from quantum to classical
physics \cite{legget}. From the theoretical point of view, spin tunneling
has been treated mainly by the use of a WKB method adapted to spin systems 
\cite{vansutto,scharf,vanwrezinski}, by using Feyman's path integral
treatment of quantum mechanics \cite{enz,schilling}, and also by using $%
su(2) $ coherent states \cite{perelomov} in order to establish a
correspondence between the spectrum of the spin system with the energy
levels of a particle moving in an effective potential \cite{ulyanov}.

In more recent years a renewed interest in the study of spin tunneling has
emerged mainly motivated by the discovery of magnetic molecules that can
stand for a decisive testing ground for the basic ideas proposed before.
From the experimental point of view, among the several advantages the
magnetic clusters present, it has been recognized that they are well defined
crystalline materials with the same shape, size and orientation.
Furthermore, the determination of their fundamental parameters which play an
essential role in the study of the dynamics of the magnetization, such as
the spin of the cluster, its magnetic anisotropy, the intra- and
inter-clusters interactions can be accurately carried out. In this
connection, the discovery of the $Mn12ac$-manganese acetate \cite{lyz} paved
the way for a series of works discussing the possibility of identifying a
spin tunneling process \cite{can,sessoli,novak,paulsen,friedman,thomas} at
sufficiently low temperatures such that the pure quantum contributions
become important. Due to the great uniaxial magnetic anisotropy of the
cluster, the first approximation phenomenological Hamiltonian for describing
this molecule is 
\bd
H=-D\mathbf{J}_{z}^{2},
\ed
where $D/k_{B}\simeq 0.7\;K$ and $k_{B}$ is the Boltzmann constant. Since it
is immediately seen that it does not contain terms responsible for
transitions between the energy levels, this Hamiltonian is not sufficient to
afford for a sound starting point for the desired study of the spin
tunneling process if only pure quantum contributions are to be considered.
Of course the thermally-assisted transitions can always be taken into
account and their role in the spin tunneling has also been emphasized as an
important component in the whole process \cite{friedman}. Along this line
sophisticated theoretical approaches have been put forth so as to treat
these important processes when temperature plays a dominant role \cite
{chudteja,leuemberger,luis}. In this way, in what concerns this molecule,
the introduction of additional terms in the Hamiltonian has already been
discussed in the literature \cite{barra2}, where the symmetry of the cluster
only allows a small transverse anisotropy of fourth-order. A more complete
Hamiltonian is then 
\bd
H=-D\mathbf{J}_{z}^{2}+B\mathbf{J}_{z}^{4}+C\left( \mathbf{J}_{+}^{4}+
\mathbf{J}_{-}^{4}\right) ,
\ed
where $B/k_{B}=-1.1\times 10^{-3}\;K$ and $C/k_{B}=\pm 3\times 10^{-5}\;K$.

The discovery of the $Fe8$ magnetic cluster (octanuclear iron cluster) \cite
{wieg} pointed to a new scenario in the spin tunneling discussions since its
proposed basic phenomenological Hamiltonian, 
\bd
H=-D\mathbf{J}_{z}^{2}+\frac{E}{2}\left( \mathbf{J}_{+}^{2}+\mathbf{J}_{-}^{2}\right) ,
\ed
with $D/k_{B}=0.275\;K$ and $E/k_{B}=0.046\;K$, already can account for a
quantum description whose results can be compared with the experimental
ones. In fact the quantum effects in the dynamics of the magnetization can
be better investigated in this magnetic cluster mainly because the
parameters characterizing it were carefully measured, and also because the
presence of the important transverse term is clearly seem to induce
tunneling effects. Also, it has been reported that this cluster has an
experimentally observed barrier of ca. $22-24$ K, and that below $0.35$
Kelvin the relaxation of magnetization becomes temperature independent \cite
{barra,sangregorio}, thus suggesting that, under this circunstance, this
temperature corresponds to the crossover to the regime where quantum effects
dominate the spin tunneling processes. Furthermore, it was also verified
that when an external magnetic field is applied along the easy axis of the $%
Fe8$ cluster the hysteresis loops present well-defined steps at integer
multiples of $\Delta H_{\parallel }\approx 0.22\;T$\cite
{barra,sangregorio,caneschi}; this also suggests, in the same way that
occurs in the $Mn12ac$ cluster, that an energy matching of states then takes
place \cite{barra,sangregorio,caneschi}. In this perspective, the $Fe8$
magnetic cluster is a good candidate to be studied in order to verify the
importance of the pure quantum contributions to the spin tunneling when
temperature effects can be neglected.

In a previous paper \cite{ga1} one of the authors has presented an
angle-based approach which allows an effective Hamiltonian to be obtained
that can be reliably used in the description of general quantum spin
tunneling processes, and in particular it was also shown that it can be used
to describe the magnetic molecules, for instance the $Fe8$ cluster. In this
case, the numerical diagonalization of the corresponding effective
Hamiltonian associated with the $Fe8$ cluster leads us to get the energy
spectrum and the ground state energy barrier in a direct way. Since the
value of the calculated energy barrier is in good agreement with the
experimental one, we want now to start from the proposed effective
Hamiltonian to get other results related to measurable properties of that
magnetic molecule.\emph{\ }In this connection, we will be interested mainly
in the ground state\emph{\ }and barrier energies, as well as in the energy
splitting that is directly associated with the spin tunneling. However,
instead of directly obtaining those results by a numerical calculation, here
we intend to show that some approximations based on well established quantum
grounds can be used that allow us to express them in analytic form. We then
propose a direct use of an adapted version of the WKB approximation \cite
{landau} for the calculation of the energy splitting of low-lying energy
levels, as well as we can also verify that the energy splitting can be
obtained for energy levels at the top of the potential barrier if, in this
case, use is made of the Kemble, Hill-Wheeler, and Miller-Good expression 
\cite{kemble,hill-wheeler,miller} for the barrier tunneling probability.

This paper is organized as follows. In section 2 we present our previous
results concerning the effective Hamiltonian related to the magnetic
molecules and some basic assumptions. In section 3 we apply these
assumptions to the case of the $Fe8$ magnetic cluster, and we compare the
results obtained with the ones coming from the direct diagonalization of the
phenomenological starting Hamiltonian of the model. The effects of the
presence of an external magnetic field oriented paralel to the easy axis on
the energy splitting is also discussed. Finally, section 4 contains our
final comments and conclusions.

\section{ Effective Hamiltonians for spin tunneling}

The starting point of the proposed approach, as discussed in Refs. \cite
{ga1,garu,gapi}, is the introduction of a quantum phenomenological
Hamiltonian describing the spin system, written in terms of angular momentum
operators obeying the standard commutation relations, being that its form
reflects the structural symmetries of the system. It may also contain terms
taking into account external applied magnetic fields. The degree of freedom
that undergoes tunneling is considered a particular collective manifestation
of the system, and it is assumed to be the only relevant one. At the same
time, the temperature of the system is assumed so conveniently low that
possible related termally assisted processes are not taken into account so
that only quantum effects are considered. For instance, the general quantum
Hamiltonian 
\be
\lb{eq1}
H=AH_{\parallel }\mathbf{J}_{z}-D\mathbf{J}_{z}^{2}+\frac{E}{2}\left( 
\mathbf{J}_{+}^{2}+\mathbf{J}_{-}^{2}\right)
\ee
can be used to study some systems of interest. In particular, for $A=g\mu _{B%
}$ this Hamiltonian describes the octanuclear iron cluster, $Fe8$, in
the presence of an external magnetic field along the $z$ axis (easy axis);
clearly, $A=0$ denotes the absence of an external field. This spin system
has a $j=S=10$ ground state and a suggested dominant quantum spin tunneling
below $0.35\;K$; furthermore, $D/k_{B}=0.275\;K$ and $E/k_{B}=0.046\;K$ \ 
\cite{barra,caneschi}. On the other hand, from a pure algebraic model point
of view, it is interesting to see that the Lipkin quasi-spin model
Hamiltonian \cite{lipkin}, of wide use in many-body physics, can also be
obtained by just considering $D=0$, being that the interest in this model
resides in the fact that it stands for a valuable testing ground for
checking the validity of approximations in treating collective degrees of
freedom.

Now, it has been already shown \cite{ga1,gapi} that a new approximate
Hamiltonian -- written in terms of an angle variable -- can be extracted
from (\ref{eq1}) which allows for a good description of spin systems when $%
2S\gtrsim 10$, where $S$ is the total spin value of the physical system. In
its general form, the new effective Hamiltonian for such a system is written
as 
\bd
H=-\frac{1}{2}\frac{d}{d\phi }\frac{1}{M\left( \phi \right) }\frac{d}{d\phi }%
+V\left( \phi \right) ,  
\ed
where
\be
\lb{eq2}
V\left( \phi \right) =V_{1}\cos ^{2}\phi +V_{2}H_{\parallel }\cos \phi +V_{3}
\ee
is the potential energy function with $V_{1}=-\left( D-E\right) S\left(
S+1\right) $, \newline
$V_{2}=-A\sqrt{S\left( S+1\right) }$ and $V_{3}=-ES\left( S+1\right) $. The
effective ``inertia'' associated with the spin system is given by 
\be
\lb{eq3}
\frac{1}{M\left( \phi \right) }=M_{1}\cos ^{2}\phi +M_{2}H_{\parallel }\cos
\phi +M_{3} , 
\ee
where $M_{1}=2\left( D-E\right) $, $M_{2}=A/S$ and $M_{3}=4E$. Figure 1
depicts the potential and effective mass functions when $A=0$ for the $Fe8$
cluster. It is clearly seen two deep minima in the potential, characterizing
the trapping wells, as well as the maxima. At the same time, we observe that
the effective mass function presents a similar behaviour, being that the
minima of both functions occur at $\phi =0$ and $\pi $.
%%%%%%%%%%%%%%%%%%%%%%%%%%%%%%%%%%%%%%%%%%%%%%%%%%%%%%%%%%%%%%%%%%%%%%%%%%%%%%%%%%%%%%%%%%%%%%%%%%%%%%%%%%%%%%%%%%%%%%%%%
\begin{figure}[!h]
\centering
\includegraphics[width=13cm]{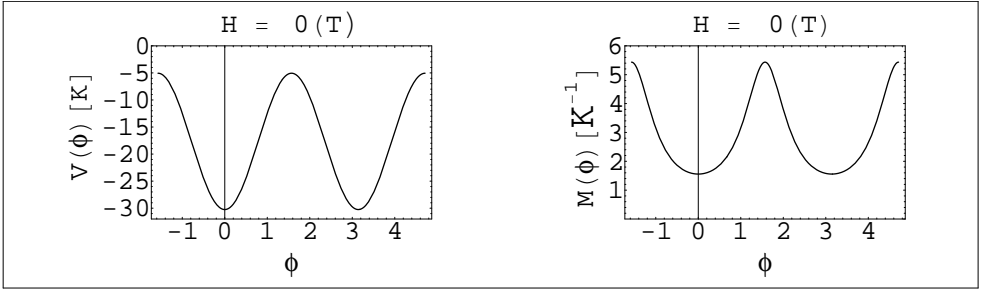}
\caption{The potential and the effective mass functions
for the $Fe8$ cluster in the absence of an external magnetic field in the
direction of the easy axis, $H_{\parallel }=0$.} 
\end{figure}
%%%%%%%%%%%%%%%%%%%%%%%%%%%%%%%%%%%%%%%%%%%%%%%%%%%%%%%%%%%%%%%%%%%%%%%%%%%%%%%%%%%%%%%%%%%%%%%%%%%%%%%%%%%%%%%%%%%%%%%%%

The corresponding Schr\"{o}dinger equation associated with the $Fe8$
molecule 
\be
\lb{eq4}
H\psi _{k}\left( \phi \right) =\mathcal{E}_{k}\psi _{k}\left( \phi \right)
\ee
can then be directly solved by just performing a Fourier analysis so that
the resulting wave function 
\bd
\psi _{k}\left( \phi \right) =\sum_{n}c_{kn}e^{in\phi } 
\ed
and energy eigenvalues are readily obtained. A test can then be immediately
carried out by comparing the eigenvalues $\mathcal{E}_{k}$ coming from the
solutions of Eq. (\ref{eq4}) with those obtained by diagonalizing Eq. (\ref
{eq1}) within the set of eigenstates $\left\{ |Sm\rangle \right\} $ of the $%
\mathbf{J}_{z}$ operator -- the results coming from the diagonalisation of
Eq. (\ref{eq1}) will be hereafter called the reference values. As already
verified in the past, the spectra in both cases are in good agreement for $%
2S\gtrsim 10$ \cite{ga1}, being the deviation of the order of $0.5\%$ for
the ground state energy and $14.7\%$ for the energy splitting of the lowest
doublet. At the same time, we can also estimate the height of the potential
barrier: since we have obtained numerically $\mathcal{E}_{gs}\simeq
-27.6447\;K$ from Eq. (\ref{eq4}), and verified that the top of the
potential barrier is given by 
\bd
V_{\max }\left( \phi \right) =-ES(S+1),  
\ed
then 
\bd
h_{b}=-ES(S+1)-\mathcal{E}_{gs}
\ed
measures the ground state energy barrier that gives $h_{b}\simeq 22.58\;K$
which is only $1.7\%$ higher than the experimental result, namely, $22.2\;K$%
, as presented in \cite{barra}.

These results then allow us to apply our approach in the description of spin
tunneling in a reliable way. Here, however, instead of focusing our
attention on the numerical procedures and results directly obtained from Eq.
(\ref{eq4}), we want to show that some assumptions based on well-established
quantum grounds can be taken for granted so that we can get analytic
expressions for the ground state and barrier energy, as well as the energy
splitting associated with spin tunneling.

\section{Application}

\subsection{Ground state and barrier energies ($H_{\parallel }=0$)}

Let us first consider the case when no external paralel magnetic field is
applied, i.e., $A=0.$ In this case we see that a simple approximate analytic
expression of the ground state energy barrier for the $Fe8$ cluster can be
obtained from the proposed effective Hamiltonian, Eqs. (\ref{eq2}) and (\ref
{eq3}). To this end, we first take into account the fact that the potential
function presents deep minima at $\phi =0$ and $\pi $, and maxima at $\ \phi
=\pi /2$ and $3\pi /2$, respectively, so that 
\bd
\mathcal{E}_{\min }=V\left( \phi =0\right) =-DS\left( S+1\right)
=-30.25\;K.\;  
\ed
We can now consider that the lowest part of the spectrum can be well
described by a harmonic approximation so that the ground state energy can be
written as $\mathcal{E}_{gs}\simeq \mathcal{E}_{\min }+\omega /2$. To obtain
an approximate expression for $\omega $ we have to consider the potential
minimum at $\phi _{\min }=0$, and verify that the effective mass can also be
taken at that angle as a good approximation in order to write 
\bd
M\left( \phi =\phi _{\min }\right) \omega ^{2}=\frac{d^{2}V\left( \phi
\right) }{d\phi ^{2}}|_{\phi _{\min }},\;  
\ed
from which we obtain 
\bd
\omega =2\sqrt{\left( D^{2}-E^{2}\right) S\left( S+1\right) } .  
\ed
It is then direct to see that the ground state energy in this approximation
is 
\bd
\mathcal{E}_{gs}=-DS\left( S+1\right) +\sqrt{\left( D^{2}-E^{2}\right)
S\left( S+1\right) }\simeq -27.41\;K ,
\ed
and, recalling that $h_{b}=-ES(S+1)-\mathcal{E}_{gs}$, the barrier energy is
written as 
\bd
h_{b}=\left( D-E\right) S\left( S+1\right) -\sqrt{\left( D^{2}-E^{2}\right)
S\left( S+1\right) }\simeq 22.35\;K\ .  
\ed
The deviation from the reference value for the ground state energy is then
of the order of $0.5\%$ while for the barrier height it is $0.7\%$. It is
interesting to verify that the analytic expression for the barrier height
can give a better approximation than the numerical calculations do.
Furthermore, since we are considering two separate potentials, tunneling is
not taken into account, but we still get good approximations to those
quantities because the change in energy due to tunneling is several orders
of magnitude smaller than the energy values of the spin states.

\subsection{Energy splitting ($H_{\parallel }=0)$}

In what concerns the energy splitting associated with tunneling, let us
assume from the outset the validity of the main ideas supporting the WKB
approximation for the present case\cite{vansutto,scharf,vanwrezinski}.
We will consider this without a formal
demonstration, but in what follows we intend to show that it is a sound
starting working background. From this standpoint, if we take advantage of
the symmetric character of both the potential function and effective
``inertia'' when $A=0$, we may go one step further and consider the WKB
energy splitting expression \cite{landau} 
\be
\lb{eq5}
\Delta \mathcal{E}\simeq \frac{\omega _{b}}{\pi }\exp \left[ -\sqrt{2%
\mathcal{M}}\int_{\phi _{i}}^{\phi _{s}}\sqrt{V\left( \phi \right) -\mathcal{%
E}}\;d\phi \right] ,
\ee
where $\omega _{b}$ is the frequency associated with the minimum of the
symmetric potential and $\phi _{i,s}$ are the classical turning points
respectively. It is important to stress that this approximation is tailored
to be used in calculating splittings in energy levels far from the top of
the potential barrier. Now, from the operational point of view we have to
observe that the mass $\mathcal{M}$ appearing in Eq. (\ref{eq5}) is
constant, whereas we have an angle-dependent effective mass function
instead. Therefore, if we want to use the WKB approximation, we have to
introduce a conveniently chosen averaging procedure in such a way that the
effective mass is then substituted by an appropriate constant value.

In order to implement all the approximations let us first consider that we
may use the approximate average value for the constant mass
\be
\lb{eq6}
\mathcal{M=}\frac{\int_{\phi _{i}}^{\phi _{s}}M\left( \phi \right) d\phi }{%
\int_{\phi _{i}}^{\phi _{s}}d\phi }, 
\ee
in the barrier region while, on the other hand, we will assume the frequency
at the minimum of the potential as before. Based on these results, we will
use the proposed approximations for the calculation of the energy splittings
associated with tunneling occuring in the energy spectrum of the $Fe8$
cluster.

First let us consider the lowest energy doublet since we are considering
that the temperature regime is such that the quantum effects dominate, i.e.,
the temperatures are below the cross-over value. Since the energy levels are
far from the barrier top, we will directly use the WKB expression, Eq. (\ref
{eq5}), where we take explicitly 
\bd
\omega _{b}=2\sqrt{\left( D^{2}-E^{2}\right) S\left( S+1\right) }
\ed
at the $\phi =0$ minimum as before, and we assume expression (\ref{eq6}) to
be valid here too. With these considerations we get $\Delta \mathcal{E}$ $%
\simeq 8.9\times 10^{-10}K$, while the reference value obtained is $\Delta 
\mathcal{E}_{ref}$ $\simeq 6.8\times 10^{-10}K$, thus showing that, in spite
of the introduced approximations, the WKB approximation is a useful
expression to estimate the energy splitting for the $Fe8$ cluster. It is
important to stress that the numerical precision plays an essential role in
the study \ of the energy splitting associated with tunneling. However,
instead of focusing estrictly on our results, although they were obtained
from double precision calculations,we will study their deviations from the
reference values since they measure the quality of our approach. Now,
considering that the next two energy doublets are also far from the barrier
top, we can go one step further and apply the WKB approximation to them.
Using the expressions for $\omega _{b}$ and $\mathcal{M}$ we get the results 
for the deviations  
\bd
\frame{%
\begin{tabular}{lll}
Doublets & Diagonalization method & WKB approximation \\ 
\textit{first} & $31\%$ & $14.7\%$ \\ 
\textit{second} & $8\%$ & $9\%$ \\ 
\textit{third} & $5\%$ & $8\%$%
\end{tabular}
}.
\ed
It is clearly seen that the obtained deviations are higher in the lowest
energy levels due to the precision of the calculations and approximations
involved.

On the other hand, if we also consider that the higher energy states can be
of importance for the transitions of the magnetic moment, we can directly
use the Kemble \cite{kemble}, Hill-Wheeler \cite{hill-wheeler}, and
Miller-Good \cite{miller} (KHW/MG) expression for the barrier penetrability
to describe the energy splitting occurring in those states that are just at
the top of the barrier, that is, 
\bd
P\left( \mathcal{E}\right) =\frac{1}{1+\exp \left[ \frac{\omega _{t}}{\pi }%
\left( V_{0}-\mathcal{E}\right) \right] },  
\ed
where $\omega _{t}$ denotes the frequency associated with the concavity at
the top of the barrier, and $V_{0}=V\left( \phi =\phi _{top}\right) $. It is
important to see that $\mathcal{E}=V_{0}$ implies in $P=0.5$ instead of the
usual WKB value $P=1.0$, thus assuring us that the KHW/MG expression is a
better approximation for this particular kind of situation.

Now, a first approximation for $\omega _{t}$ can be proposed as in the
previous case, i.e., 
\bd
\omega _{t}=\sqrt{\frac{1}{M\left( \phi =\phi _{top}\right) }\left| \frac{%
d^{2}V\left( \phi \right) }{d\phi ^{2}}|_{\phi =\phi _{top}}\right| }.
\ed
Since the barrier presents a maximum at $\phi =\pi /2$ where $V_{0}=-5.06\;K$%
, as already seen, and the reference results show that there is a doublet in
this energy region, then we are allowed, in the same way we did before, to
use the KHW/MG approximation to estimate the corresponding energy splitting.
Thus, at the barrier top the frequency is given by 
\bd
\omega _{t}=2\sqrt{2E\left( D-E\right) S\left( S+1\right) }, 
\ed
and the effective mass at the maximum is 
\bd
M\left( \phi =\frac{\pi }{2}\right) =\frac{1}{4E}\ ,
\ed
so that 
\bd
\Delta \mathcal{E\simeq }\frac{2\sqrt{\left( D^{2}-E^{2}\right) S\left(
S+1\right) }}{\pi }\left\{ 1+\exp \left[ \frac{\pi \left( \left| \mathcal{E}%
\right| -ES\left( S+1\right) \right) }{\sqrt{2E\left( D-E\right) S\left(
S+1\right) }}\right] \right\} ^{-1}  
\ed
is an estimate to the energy splitting at the barrier top. In this way, for
the energy levels at $\left| \mathcal{E}\right| \simeq 5.34\;K$ we get $%
\Delta \mathcal{E\simeq }0.65\;K$, being the reference result $\Delta 
\mathcal{E}_{ref}\simeq 0.72\;K$. These results indicate that the KHW/MG
approximation can be trustfully used in estimating the energy splitting at
the top of the $Fe8$ cluster energy barrier. However, its use for the next
lower doublet already presents a great deviation from the reference value.
For those doublets close to (but not at) the barrier top, a parabolic
approximation is much more adequate. In this case, the energy splitting
expression takes the form 
\bd
\Delta \mathcal{E}_{par}\mathcal{\simeq }\frac{\omega _{b}}{\pi }\exp \left[
-\frac{\pi \left( \left| \mathcal{E}\right| -V_{0}\right) }{2\sqrt{2Eh_{p}}}%
\right] ,  
\ed
where 
\bd
h_{p}=V\left( \phi _{\max }\right) -V\left( \phi _{\min }\right)
\ed
is the total potential energy barrier depth. Applying this expression to the
levels near $\left| \mathcal{E}\right| \simeq 7.5\;K$ we get $\Delta 
\mathcal{E}_{par}\simeq 0.14\;K$, whereas $\Delta \mathcal{E}_{ref}\simeq
0.13\;K$.

\subsection{Low-lying Energies and Energy splitting ($H_{\parallel }\neq 0$)}

We can discuss now the role of an applied external magnetic field along the $%
z$ direction (easy axis) on the $Fe8$ cluster. In this case, if such an
external magnetic field is applied we start from Eqs. (\ref{eq2}) and (\ref
{eq3}) with $A=g\mu _{B}$, with fixed $g=2$.

It is immediate to see that the presence of the magnetic field does not
change the positions of the potential extrema at $\phi =0$ and $\pi $, but
it introduces a shift in energy at these points so that the difference in
the height between the two minima will be $\left( H_{\parallel }>0\right) $%
\bd
V\left( \phi =\pi \right) -V\left( \phi =0\right) =2\sqrt{S\left( S+1\right) 
}g\mu _{B}H_{\parallel }\ .
\ed
This means that for some particular values of $H_{\parallel }$ a degeneracy
in the energy spectrum will occur such that the state with quantum number,
for instance, $n$, will match its energy with that of the state with $-n+k$,
as it is indeed expected. Figure 2 exhibits the potential and effective mass
for some values of $H_{\parallel }$. The pictures depict the effects of the
external field on the potencial and effective mass functions, both for a
weak field (that produces the first matching) as well as for a value near
that producing an diverging behaviour of the effective mass.
Let us first discuss weak external magnetic $H_{\parallel }$ fields, below
the first matching value. Following the same procedure developed above,
where use was made of a harmonic approximation, we see that the energy of
the ground state in the deeper/shallower potential well is given by 
\bd
\mathcal{E}_{gs}(H_{\parallel })\simeq \mathcal{E}_{gs}(H_{\parallel }=0)\pm
\ed
\bd
\left\{ V_{2}+\frac{1}{2}\sqrt{-2V_{1}\left( M_{1}+M_{3}\right) }\left[ 
\frac{V_{2}}{4V_{1}}+\frac{M_{2}}{2\left( M_{1}+M_{3}\right) }\right]
\right\} H_{\parallel }\ .
\ed
%
%%%%%%%%%%%%%%%%%%%%%%%%%%%%%%%%%%%%%%%%%%%%%%%%%%%%%%%%%%%%%%%%%%%%%%%%%%%%%%%%%%%%%%%%%%%%%%%%%%%%%%%%%%%%%%%%%%%%%%%%%
\begin{figure}[!h]
\centering
\includegraphics[width=13cm]{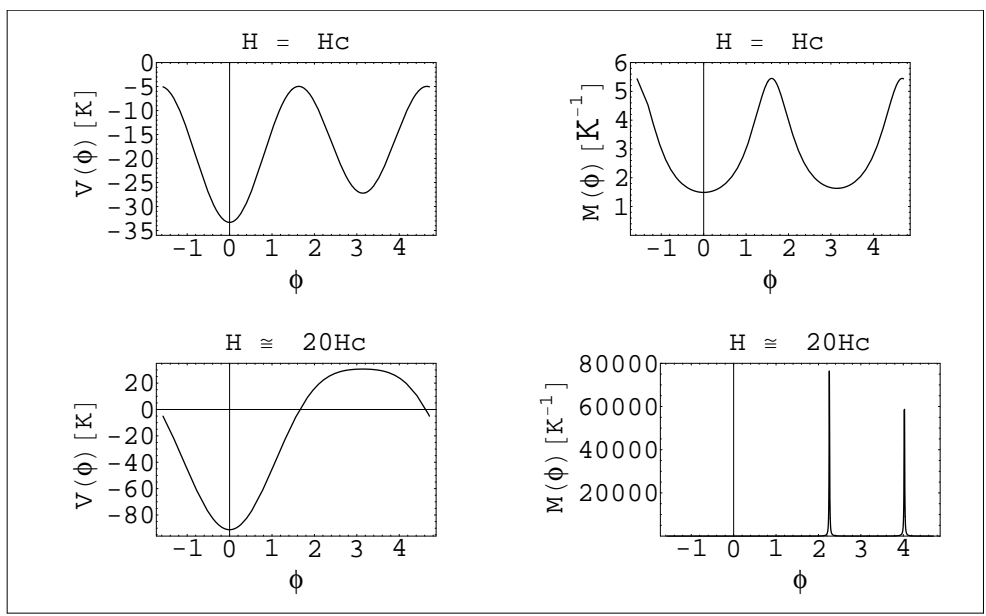}
\caption{The potential and the effective mass functions
for the $Fe8$ cluster for some values of the external magnetic field in the
direction of the easy axis. In case (a) the magnetic field corresponds to
that of the first matching of the energy levels, while, in case (b), the
magnetic field is close to the saturation value. It is to be noted that the
effective mass in case (b) exhibits values already showing the trend to
divergence that occurs when the magnetic field attains the saturation value.
} 
\end{figure}
%%%%%%%%%%%%%%%%%%%%%%%%%%%%%%%%%%%%%%%%%%%%%%%%%%%%%%%%%%%%%%%%%%%%%%%%%%%%%%%%%%%%%%%%%%%%%%%%%%%%%%%%%%%%%%%%%%%%%%%%%

It is important to see that in this approximation, since we are considering
two separate potentials, there is no tunneling at all, so that the two
levels coincide for $H_{\parallel }=0$. Therefore, the energy gap as a
function of the external magnetic field $H_{\parallel }$ can be directly
obtained, and reads 
\bd
\Delta \mathcal{E}\left( H_{\parallel }\right) \simeq \Delta \mathcal{E}%
\left( H_{\parallel }=0\right) +
\ed
\be
\lb{eq7}
\left\{ 2V_{2}+\sqrt{-2V_{1}\left( M_{1}+M_{3}\right) }\left[ \frac{V_{2}}{%
4V_{1}}+\frac{M_{2}}{2\left( M_{1}+M_{3}\right) }\right] \right\}
H_{\parallel }  
\ee
if we also take into account now the tunneling. This result can be
immediately compared with the reference values obtained from Eq. (\ref{eq1}%
). Our result for the coefficient of the linear dependence is $%
26.79\;KT^{-1} $, while the reference result, which also exhibits a linear
dependence, is $26.85\;KT^{-1}$, thus showing a good agreement with a
deviation of $0.22\%$.

In what concerns specifically the discussion of the change of the energy
splitting associated with the tunneling due to the presence of an external
magnetic field of this kind, we have to point that a new expression, still
based on the WKB approximation, must be introduced because of the
field-induced asymmetry of the potential and effective mass (see figure 2).
It can be shown that the new expression is given by 
\bd
\Delta \mathcal{E}\left( H_{\parallel }\right) \simeq \frac{\sqrt{\omega
_{1}\omega _{2}}}{\pi }\exp \left( -\sqrt{2\mathcal{M}}\int_{\phi
_{i}}^{\phi _{s}}\sqrt{V_{1}\cos ^{2}\phi +V_{2}H_{\parallel }\cos \phi
+\left| \mathcal{E}\right| +V_{3}}\;d\phi \right) ,
\ed
where $\omega _{1,2}\left( H_{\parallel }\right) $ are the frequencies at
the minima of the two potential wells and the effective mass can be treated
in the same way as before. If we perform a harmonic approximation for both $%
\omega _{1}$ and $\omega _{2}$ we can readily conclude that related factor
is nearly constant for weak $H_{\parallel }$. In the same way, an analysis
of the behaviour of the effective mass as a function of small $H_{\parallel
} $ also shows that it is also almost constant. In fact the dominant
contribution comes from the exponential factor -- the tunneling probability
factor -- which can be given in an approximated form as 
\bd
\exp \left[ -\sqrt{2\mathcal{M}}\int_{\phi _{i}}^{\phi _{s}}\sqrt{V_{1}\cos
^{2}\phi +V_{2}H_{\parallel }\cos \phi +\left| \mathcal{E}\right| +V_{3}}%
\;d\phi \right] \simeq
\ed
\be
\lb{eq8}
\exp \left[ \left( -\sqrt{2\mathcal{M}}\int_{\varphi _{i}}^{\varphi _{s}}%
\sqrt{V_{1}\cos ^{2}\phi +\left| \mathcal{E}\right| +V_{3}}\;d\phi \right)
\left( 1+\chi H_{\parallel }^{2}\right) \right] \ ,
\ee
where $\chi $ is a constant and the integral in the rhs corresponds to the
tunneling probability factor in the absence of the magnetic field. Therefore
we see that the total energy separation of doublets tends to increase
linearly (as can be seen from Eq. (\ref{eq7})) with small $H_{\parallel }$
-- thus removing degeneracies that would occur when $H_{\parallel }=0$ if no
tunneling is considered -- because of the asymmetry of the potential wells,
at the same time that the tunneling probability factor tends to decrease
with small $H_{\parallel }$ mainly due to the widening of the barrier to be
tunneled, Eq. (\ref{eq8}).

Another interesting result that can be still obtained with the harmonic
approximations is the numerical value of the external magnetic field that
result in the matching of the energy levels in the two separate potential
wells. In fact, by imposing that the first excited state in the deeper
potential must match the lowest energy level in the shallower one, we get an
expression that gives that particular value of the external magnetic field.
Therefore, since we already know the expressions for $\mathcal{E}_{\min }$
and $\omega $, we write 
\bd
\mathcal{E}_{first}^{deep}\left( H_{\parallel }\right) \simeq \mathcal{E}%
_{\min }\left( H_{\parallel }\right) +\frac{3}{2}\omega \left( H_{\parallel
}\right)
\ed
and

\bd
\mathcal{E}_{gs}^{shallow}\left( H_{\parallel }\right) \simeq \mathcal{E}%
_{\min }\left( H_{\parallel }\right) +\frac{1}{2}\omega \left( H_{\parallel
}\right) \ .
\ed
Now, imposing that the two levels match their energies, we get 
\bd
H_{\parallel }^{0}=\frac{-\sqrt{-2V_{1}\left( M_{1}+M_{3}\right) }}{2\left\{ 
\sqrt{-2V_{1}\left( M_{1}+M_{3}\right) }\left[ \frac{V_{2}}{4V_{1}}+\frac{%
M_{2}}{2\left( M_{1}+M_{3}\right) }\right] +V_{2}\right\} }\ .
\ed
The estimated magnitude of the field is then $H_{\parallel }^{0}\approx
0.2239\;T$ which is in good agreement with the experimental value of $%
0.22\;T $ as presented in \cite{barra,caneschi}.

From another perspective, it is interesting to observe that the particular
value of the magnetic field intensity $H_{0}$, whose multiples $kH_{0}$ lead
to the matching of the energy levels, can be obtained without any reference
to the harmonic approximation at all. In fact, we can also extract an
expression for that particular value of the magnetic field from an analysis
of the effective mass expression. Realizing that the presence of zeroes in
the function $I\left( \phi \right) =1/M\left( \phi \right) $ (infinities of
the effective mass) indicates that tunneling cannot occur, we look for the
expression of the strength of $H_{\parallel }$ beyond which tunneling will
not take place. A direct calculation shows that the limit - the saturation
value - is given by

\be
\lb{eq9}
H_{\parallel }^{\lim }=\frac{4Sk_{B}}{g\mu _{B}}\sqrt{2E\left( D-E\right) }%
\simeq 4.32\;T  
\ee
so that 
\be
\lb{eq10}
H_{\parallel }^{0}=\frac{H_{\parallel }^{\lim }}{2S}=\frac{2k_{B}}{g\mu _{B}}%
\sqrt{2E\left( D-E\right) }\simeq 0.216\;T  
\ee
for $g=2$. This result is also in good agreement with the experimental
value. It is important to verify that for this value of the external paralel
magnetic field the original minimum at $\phi =\pi $ and the maxima at $\phi
=\pi /2$ and $3\pi /2$ have turned into a single maximum of the potential
function, while the only surviving minimum is the one at $\phi =0\;(=2\pi )$%
. This means that, in this particular situation, there is only one direction
along which the spin can be directed at.

\section{Conclusions}

Based on a quantum angle-based description of spin tunneling developed
before and briefly reviewed here, in this paper we addressed the problem of
obtaining quantitative results for the $Fe8$ cluster in what concerns the
barrier, low-lying energy doublets and their energy splittings associated
with spin tunneling. Although numerical calculations are direct, and the
results are in very good agreement with those coming from the
diagonalization of the starting phenomenological Hamiltonian -- written in
terms of the angular momentum operators --, we have shown that analytic
expressions can be obtained that allow us to see how the quantities of
interest of the $Fe8$ cluster depend on its anisotropy constants. This
construction can be seen to be very simple in the case of no external
magnetic field, when a harmonic approximation for the potential energy
function around its minimum, together with the introduction of the
corresponding frequency at that point, is proposed. Analytic expressions for
the ground state and barrier energies are then directly written. In what
regards the energy splitting associated with the spin tunneling in this
case, we have shown that an adapted WKB expression for symmetric potentials
can be directly used and that the results are of the same order of those
coming from the numerical calculations. Thus, we verify that the WKB
expression can be reliably used in the description of tunneling occuring in
the $Fe8$ cluster.

We also have shown that we can still obtain the relevant results if an
external magnetic field paralel to the easy axis is applied. In this case,
besides the expressions for the low-lying energy spectrum, we also obtained
an analytic expression for the rule governing the spacing between the energy
doublets as a function of the intensity of the magnetic field. The linear
dependence thus obtained can be seen to come from the asymmetry of the
potentials induced by the external magnetic field. It is also clear that
this effect is several orders of magnitude higher than the energy splitting
associated with tunneling. This quantum effect, by its turn, can also be
described in the case of a slightly asymmetric potential through an extended
WKB expression, and it was shown that the energy splitting increases with
the intensity of the external magnetic field, as it should be, mainly due to
the widening of the potential barrier.

Experimentally, the value of the external magnetic field whose multiples
produce well-defined steps in the hysteresis loops was determined, and it
was also suggested that such effect corresponds to the energy matching of
states in the two wells of an energy barrier. In the present quantum
angle-based approach, we have shown that this is a consistent picture of the
process, and that the experimental value $H_{\parallel }^{0}=0.22\;T$ is
obtained through a simple approximation proposed to describe the low-lying
energy levels of the spectrum of the $Fe8$ cluster. In fact, we also have
shown that the result thus obtained is not fortuitous; regarding the
expression for the effective mass as another valuable source of information
about the tunneling process, we obtained almost the same value for $%
H_{\parallel }$ by simply looking for the intensity of the field at which
the effective mass diverges, thus blocking the tunneling process.

The agreement between the results presented in this paper and the
experimental data concerning the $Fe8$ cluster properties seems to
corroborate the usefulness of our quantum angle-based proposed approach
indicating that it can be reliably used in the description of tunneling of
spin systems.
%
%%%%%%%%%%%%%%%%%%%%%%%%%%%%%%%%%%%%%%%%%%%%%%%%%%%%%%%%%%%%%%%%%%%%%%%%%%%%%%%%%%%%%%%%%%%%%%%%%%%%%%%%%%%%%%%%%%%%%%%%%
\ack 
%%%%%%%%%%%%%%%%%%%%%%%%%%%%%%%%%%%%%%%%%%%%%%%%%%%%%%%%%%%%%%%%%%%%%%%%%%%%%%%%%%%%%%%%%%%%%%%%%%%%%%%%%%%%%%%%%%%%%%%%%
The authors would like to thank Dr. M. Marchiolli from Instituto de F\'{i}%
sica Te\'{o}rica - UNESP for a careful reading of the manuscript. D.G. was
partially supported by Conselho Nacional de Desenvolvimento Cient\'{i}fico e
Tecnol\'{o}gico, CNPq, and E.C.S. was supported by CAPES.

\begin{thebibliography}{99}
\bibitem{takagi}  S. Takagi, \textit{Macroscopic Quantum Tunneling},
Cambridge University Press, Cambridge, UK, 2002.

\bibitem{chudteja}  E.M. Chudnovsky, J. Tejada, \textit{Macroscopic Quantum
Tunneling of the Magnetic Moment}, Cambridge University Press, UK, 1998.

\bibitem{legget}  A.J. Legget, S. Chakravarty, A.T. Dorsey, M.P.A. Fisher,
A. Garg, W. Zwerger, Rev. Mod. Phys. 59 (1987) 1.

\bibitem{vansutto}  J.L. van Hemmen, A. S\"{u}to, Europhys. Lett 1 (1986)
481; J.L. van Hemmen, A. S\"{u}to, Physica 141\textbf{B} (1986) 37.

\bibitem{scharf}  G. Scharf, W.F. Wreszinski, J.L. van Hemmen, J. Phys. 
\textbf{A }20 (1987) 4309.

\bibitem{vanwrezinski}  J.L. van Hemmen, W.F. Wreszinski, Comm. Math. Phys.
119 (1988) 213.

\bibitem{enz}  M. Enz, R. Schilling, J. Phys. \textbf{C }19 (1986) 1765; M.
Enz, R. Schilling, J. Phys. \textbf{C }19 (1986) L711.

\bibitem{schilling}  R. Schilling, Proc. of the NATO Advanced Reaserch
Workshop on Quantum Tunneling of Magnetization, Kluwer, Dordrecht (1995).

\bibitem{perelomov}  A. Peremolov, \textit{Generalized Coherent States and
Their Applications, }Springer Verlag, Berlin, 1986.

\bibitem{ulyanov}  V.V. Ulyanov, O.B. Zaslavskii, Phys. Rep. 216\textbf{C }%
(1992) 179.

\bibitem{lyz}  T. Lis, Acta Crystallographica \textbf{B }36 (1980) 2042.

\bibitem{can}  A. Caneschi, D. Gateschi, R. Sessoli, A.L. Barra, L.C.
Brunel, M. Guillot, J. Am. Chem. Soc. 113 (1991) 5873.

\bibitem{sessoli}  R. Sessoli, D. Gatteschi, A. Caneschi, M.A. Novak, Nature
(London) 365 (1993) 141.

\bibitem{novak}  M.A. Novak, R. Sessoli, A. Caneschi, D. Gatteschi, J. Magn.
Magn. Mater. 146 (1995) 211.

\bibitem{paulsen}  C. Paulsen, J.-G. Park, B. Barbara, R. Sessoli, A.
Caneschi, J. Magn. Magn. Mater. 140-144 (1995) 379, 1891.

\bibitem{friedman}  J.R. Friedman, M.P. Sarachik, J. Tejada, R. Ziolo, Phys.
Rev. Lett. 76 (1996) 3830.

\bibitem{thomas}  L. Thomas, F. Lionti, R. Ballou, D. Gatteschi, R. Sessoli,
B. Barbara, Lett. Nature, 383 (1996) 145.

\bibitem{leuemberger}  M.N. Leuenberger, D. Loss, Europhys. Lett. 46 (1999)
692; Phys. Rev. \textbf{B }61 (2000) 1286.

\bibitem{luis}  J.F. Fern\'{a}ndez, J. Bartolom\'{e}, F. Luis, J. of Appl.
Phys. 83 (1998) 6940; Phys. Rev. \textbf{B }57 (1998) 505.

\bibitem{barra2}  A.L. Barra, D. Gatteschi, R. Sessoli, Phys. Rev. \textbf{B 
}56 (1997) 8192.

\bibitem{wieg}  K. Wieghardt, K. Pohl, I. Jibril, G. Huttner, Angew. Chem.
Int. Ed. Engl. 23\textbf{\ }(1984) 77.

\bibitem{barra}  A.L. Barra, P. Debrunner, D. Gatteschi, Ch.E. Schulz, R.
Sessoli, Europhys. Lett. 35 (1996) 133.

\bibitem{sangregorio}  C. Sangregorio, T. Ohm, C. Paulsen, R. Sessoli, D.
Gatteschi, Phys. Rev. Lett. 78 (1997) 4645.

\bibitem{caneschi}  A. Caneschi, D, Gatteschi, C. Sangregorio, R. Sessoli,
L. Sorace, A. Cornia, M.A. Novak, C. Paulsen, W. Wernsdorfer, J. Magn. Magn.
Mater. 200\textbf{\ }(1999) 182.

\bibitem{ga1}  D. Galetti, Physica 374\textbf{A }(2007) 211.
doi:10.1016/j.physa.2006.07.030.

\bibitem{landau}  L. Landau, \textit{M\'{e}chanique Quantique}, MIR, 1989,
Chap VII.

\bibitem{kemble}  E.C. Kemble, \textit{Fundamental Principles of Quantum
Mechanics}, McGrawHill, New York, 1937.

\bibitem{hill-wheeler}  D.Hill, J.A. Wheeler, Phys. Rev. 89 (1953) 1102.

\bibitem{miller}  S.C. Miller, R.M. Good, Phys. Rev. 91 (1953) 174.

\bibitem{garu}  D. Galetti, M.Ruzzi, J. Phys. A: Math. Gen. 33 (2000) 2799.

\bibitem{gapi}  D. Galetti, B.M. Pimentel, C.L. Lima, Physica 351\textbf{A }%
(2005) 315.

\bibitem{lipkin}  H. Lipkin, N. Meshkov, A.J. Glick, Nucl. Phys. 62 (1965)
188.
\end{thebibliography}
\end{document}